\documentstyle[12pt]{article}

\begin{document}
\setcounter{page}{1}

\pagestyle{plain}
\vspace{1cm}
\begin{center}
\Large{\bf Comparison of Approaches to Quantum Correction of Black Hole Thermodynamics}\\
\small \vspace{1cm}
{\bf Kourosh Nozari}\quad and \quad{\bf A. S.  Sefidgar}\\
\vspace{0.5cm} {\it Department of Physics,
Faculty of Basic Science,\\
University of Mazandaran,\\
P. O. Box 47416-1467,
Babolsar, IRAN\\
e-mail: knozari@umz.ac.ir}
\end{center}
\vspace{1.5cm}
\begin{abstract}
There are several approaches to quantum gravitational corrections of
black hole thermodynamics. String theory and loop quantum gravity,
by direct analysis on the basis of quantum properties of black
holes, show that in the entropy-area relation the leading order
correction should be of log-area type. On the other hand,
generalized uncertainty principle(GUP) and modified dispersion
relations(MDRs) provide perturbational framework for such
modifications. Although both GUP and MDRs are common features of all
quantum gravity scenarios, their functional forms are quantum
gravity model dependent. Since both string theory and loop quantum
gravity give more reliable solution of the black hole
thermodynamics, one can use their results to test approximate
results of GUP and MDRs. In this paper, we find quantum corrected
black hole thermodynamics in the framework of GUP and MDR and then
we compare our results with string theory solutions. This comparison
suggests severe constraints on the functional form of GUP and MDRs.
These constraints may reflect characteristic features of ultimate
quantum gravity theory.
\end{abstract}
\section{Motivation}
A common feature of all promising candidates for quantum gravity is
existence of minimal observable length, which is on the order of
Planck length[1-5]. There are several approaches to incorporate this
finite resolution of spacetime with theoretical framework of
standard model. GUP and MDRs are two of these approaches. In fact,
GUP and MDRs are common features of all candidates for quantum
gravity. In particular, in the study of loop quantum gravity and of
models based on noncommutative geometry, there has been strong
interest in some candidate modifications of the energy-momentum
dispersion relations[6-10] . On the other hand, generalized
uncertainty principles have been considered primarily in the
literature on string theory and on models based on noncommutative
geometry[1-5]. Possible relations between GUP and MDRs has been
studied recently[11]. It is natural to expect that GUP and MDRs
affect black hole thermodynamics, since black hole structure is an
example of extreme quantum gravity regime. Any constraint imposed on
the form of GUP and MDRs in study of black hole physics, will help
us to find more accurate form of ultimate quantum gravity scenario.
Black holes thermodynamics in the framework of GUP and MDRs has been
studied by several authors[12-23]. Recently, Amelino-Camelia {\it et
al} have studied this issue with details[9,10]. They have argued
that for consistency between string theory results and the results
of MDRs, the term proportional to first order of Planck length in
MDRs should not be present. Here we are going to proceed further in
this direction. We will show that comparison between results of
string theory and MDRs, suggests that all terms proportional to odd
power of energy should not be present in MDRs. On the other hand,
comparison between results of string theory and GUP suggests that in
GUP even power of $\delta x$ should not be present. These two
important results restrict the form of MDRs and GUP considerably.
Naturally, this restrictions may show some characteristic features
of underlying quantum gravity theory. In addition, our comparison
between results of GUP and MDR show that these to features of
quantum gravity are not different considerably and they would be
equivalent in ultimate quantum gravity theory.\\
In which follows we set $\hbar=c=G=1$.

\section{Preliminaries}
In this section we provide some preliminaries for rest of the paper.
\subsection{MDR}
A modified Dispersion Relation(MDR) can be written as [9]
\begin{equation}
 ({\vec p})^2=f(E,m;L_P)\simeq {E}^2 - {\mu}^2 + {\alpha}_1 {L}_P
{E}^3 + {\alpha}_2 {L}_{P}^{2} E^4 + O({L}_P^3{E}^5)
\end{equation}
where $f$ is the function that gives the exact dispersion relation,
and on the right-hand side we have assumed the applicability of a
Taylor-series expansion for $E\ll \frac{1}{L_P}$ . The coefficients
$\alpha_i$ can take different values in different quantum-gravity
proposals. Note that $m$ is the rest energy of the particle and the
mass parameter $\mu$ on the right-hand side is directly related to
the rest energy, but $\mu \neq m$ if the $\alpha_i$ do not all
vanish.
\subsection{GUP}
A generalized uncertainty principle(GUP), can be written as
follows[9]
\begin{equation}
\delta x \geq \frac{1} {\delta p} + \alpha l_P^2 \delta p +O(l_P^3
\delta p^2)
\end{equation}
which has been derived within the string theory approach to the
quantum-gravity problem and several alternative scenarios. This GUP
is such that at small $\delta p$ one finds the standard dependence
of $\delta x$ on $\delta p$ ( $\delta x$ gets smaller as $\delta p$
increases ) but for large $\delta p$ the Planckian corrections term
becomes significant and keeps $\delta x \geq L_P $. Within string
theory, the coefficient $\alpha$ should take a value of roughly the
ratio between the square of the string length and the square of the
planck length , but this of course might work out differently in
other quantum-gravity proposals.
\subsection{String Theory Results for Black Hole Thermodynamics}
Bekenstein-Hawking formalism of black hole thermodynamics should be
modified to incorporate quantum gravitational effects. Both GUP and
MDRs provide a perturbational framework for these
modifications[12-23]. On the other hand, loop quantum gravity and
string theory give reliable entropy-area relation of the black holes
(for $A \gg L_P^2)$,
\begin{equation}
S=\frac{A}{4{L}_P^2}+\rho \ln {\frac {A}{L_P^2}} + O(\frac
{L_P^2}{A}),
\end{equation}
where $\rho$ might take different values in string theory and in
loop quantum gravity[9,10,24]. If we use the relation
\begin{equation}
S=\frac{A}{4{L}_P^2}+\rho \ln {\frac {A}{L_P^2}} + \beta \frac
{L_P^2}{A},
\end{equation}
we can derive the mass-temperature relation of the black holes as,
\begin{equation}
T=\frac{L_p^2}{8 \pi M} \Bigg( 1 - \rho \frac {L_p^2 }{4 \pi M^2} +
\frac {L_p^4 }{ (4 \pi)^2M^4 } (\rho^2 + \frac {\beta}{4})\Bigg).
\end{equation}
Now the question arises: are the entropies calculated within GUP and
MDR consistent with the string theory results? To answer this
question, first we should calculate entropies within GUP and MDR. In
this approach we will use the fact that when a quantum particle with
energy $E$ and size $l$ is absorbed into a black-hole and $l \sim
\delta x $, the minimum increase of area of black-hole will be
\begin{equation}
\Delta A \geq 4 (\ln 2) L_P^2 E \delta x
\end{equation}
and the minimum increase of entropy is $\ln2$, which can be
interpreted as one bit of information [25, 26].

\section{GUP and Black Hole Thermodynamics}
Consider the following GUP
\begin{equation}
\delta p \geq \frac{1}{\delta x} (1+\alpha L_P^2 \delta p^2).
\end{equation}
This relation can be written as
\begin{equation}
\delta p \geq \frac{1}{\delta x} \Big[1+\frac {\alpha L_P^2}{\delta
x^2}\Big(1+\alpha L_P^2 \delta p^2\Big)^2\Big].
\end{equation}
Considering only lowest order terms in the power of $L_{p}$, we find
\begin{equation}
\delta p \geq \frac{1}{\delta x} (1+\frac{\alpha L_P^2}{ \delta
x^2}).
\end{equation}
Using standard dispersion relation $p=E$, we find
\begin{equation}
\delta E \geq \frac{1}{\delta x} (1+\frac{\alpha L_P^2}{ \delta
x^2}).
\end{equation}
Generally this relation can be written as
\begin {equation}
E \geq \frac{1}{\delta x} + \frac{\alpha L_P^2}{ \delta x^3}+
O\Big(\frac {L_P^3}{(\delta x)^4}\Big).
\end{equation}
In their analysis, Amelino-Camelia {\it et al} have used this
relation with only two first terms of the right hand side[9,10].
Here we consider more terms to explore their effects on the black
hole entropy. When we compare our results with the standard results
of string theory, our comparison will suggests severe
constraints on the general form of GUP.\\
Consider the following generalization
\begin {equation}
E \geq \frac{1}{\delta x} + \frac{\alpha L_P^2}{ \delta x^3}+ \frac
{\alpha^{'} L_P^3}{\delta x^4} + \frac {\alpha^{''} L_p^4}{\delta
x^5} +\frac {\alpha^{'''} L_p^5}{\delta x^6},
\end{equation}
which leads to
\begin{equation}
E \delta x \geq 1 + \frac{\alpha L_P^2}{ \delta x^2}+ \frac
{\alpha^{'} L_P^3}{\delta x^3} + \frac {\alpha^{''} L_p^4}{\delta
x^4} +\frac {\alpha^{'''} L_p^5}{\delta x^5}.
\end{equation}
Substituting the minimum value of $ E\delta x $ in (6), we find,
\begin{equation}
\Delta A \geq 4 (\ln2) L_P^2\Big[1 + \frac{\alpha L_P^2}{\delta
x^2}+\frac {\alpha ^{'} L_p^3}{\delta x^3} +\frac {\alpha ^{''}
L_p^4}{\delta x^4} + \frac {\alpha^{'''} L_P^5}{\delta x^5}\Big].
\end{equation}
This relation can be written approximately as
\begin{equation}
 \frac{dS}{dA} \approx \frac {\Delta S_(min)}{\Delta A_(min)}\simeq\frac
{\ln2}{4 (\ln2) L_p^2\Big( 1 + \frac{\alpha L_P^2}{ \delta x^2}+
\frac {\alpha^{'} L_P^3}{\delta x^3} + \frac {\alpha^{''}
L_p^4}{\delta x^4} +\frac {\alpha^{'''} L_p^5}{\delta x^5}\Big)},
\end{equation}
which leads to
\begin{equation}
\frac {dS}{dA}\simeq\frac{1}{4 L_p^2} \Bigg[1 - \alpha L_p^2 \frac
{1}{\delta x^2} - \alpha^{'} L_p^3 \frac{1}{\delta x^3} + (\alpha
^{2} - \alpha^{''}) \frac{L_p^4}{\delta x^4} + ( 2\alpha \alpha^{'}
- \alpha^{'''} ) \frac {L_p^5}{\delta x^5}\Bigg ],
\end{equation}
where we have neglected terms with order higher than $O\Big(\frac
{L_p^5}{\delta x^5}\Big)$. Using $ A=4 \pi R_s^2 \simeq 4 \pi \delta
x^2$ where $R_{s}$ is radius of black hole event horizon (here we
have assumed that in falling in the black hole, the particle
acquires position uncertainty $\delta x\sim R_{s}$ [25,26]), we can
integrate to find
\begin{equation}
S \simeq \frac{A}{4 L_p^2} - \pi \alpha \ln \frac{A}{L_p^2} + 4 \pi
^{\frac {3}{2}} \alpha ^{'} L_p A ^{\frac{-1}{2}} - (\alpha^2 -
\alpha ^{''} ) L_p^2 4 \pi^2 A^{-1} - \frac{16}{3} L_p^3
\pi^{\frac{5}{2}} (2 \alpha \alpha^{'} - \alpha ^ {'''}) A ^ {\frac
{-3}{2} }.
\end{equation}
Assuming that string theory result (4), is correct, we should
conclude that $\alpha' = \alpha'''=0$. This means that in GUP (12),
all terms with even power of $\frac{1}{\delta x} $ should be
omitted. That is, only even power of Planck length cold appear in
GUP. Therefore, within GUP, black hole entropy is given by
\begin{equation}
S \simeq \frac{A}{4 L_p^2} - \pi \alpha \ln \frac{A}{L_p^2}- 4
\pi^2(\alpha^2-\alpha^{''}) \frac{L_p^2}{A}.
\end{equation}
Comparing this result with (4) suggests that $\rho=-\pi \alpha$ and
$\beta=-4\pi^{2}(\alpha^{2}-\alpha^{''})$. Since according to string
theory, $\rho$ and $\beta$ are given, then $\alpha$ and
$\alpha^{''}$ are determined and our GUP is well established. Using
the familiar relation between black hole area and mass $A = 16 \pi
M^2$ and the first law of black hole thermodynamics, $dS =\frac
{dM}{T} $, we can easily obtain the temperature of the black hole
\begin{equation}
T \simeq \frac{L_p^2}{8 \pi M} [ 1 + \frac { \alpha L_p^2}{4 M^2}  +
\frac{\alpha^{'} L_p^3}{8 M^3}+ \frac{\alpha^{''}L_p^4}{16 M^4} +
\frac{\alpha^{'''} L_p^5}{32 M^5}]
\end{equation}
Comparison between our result and string theory result (5), shows
that the coefficients of even powers of $\frac{1}{M}$ which are not
present in string theory result, should be vanishing. This leads us
to $\alpha' = \alpha'''=0$ once again. Therefore, our comparison
restricts the form of GUP to having only even power of $L_{p}$, that
is
\begin {equation}
E \geq \frac{1}{\delta x} + \frac{\alpha L_P^2}{ \delta x^3}+ \frac
{\alpha^{''} L_p^4}{\delta x^5} +\frac {\alpha^{(4)} L_p^6}{\delta
x^7}+...\quad .
\end{equation}
Since GUP is a model independent concept, any constraint on the form
of GUP (such as our finding) can be attributed to the nature of
ultimate quantum gravity theory. In other words, constraints imposed
on the form of GUP will help us to find deeper insight to the nature
of underlying quantum gravity theory.

\section{MDR and Black Hole Thermodynamics}
In this section we derive the entropy and temperature of the black
hole within MDR and the standard uncertainty principle. Then we
compare our results with standard string theory results to find more
concrete form of MDR. We use a more general form of MDR relative to
(1),
\begin{equation}
({\vec p})^2=f(E,m;L_P)\simeq {E}^2 - {\mu}^2 + {\alpha}_1 {L}_P
{E}^3 + {\alpha}_2 {L}_{P}^{2} E^4 + \alpha_3{L}_P^3{E}^5+
\alpha_{4}{L}_P^4{E}^6 +O({L}_P^5{E}^7).
\end{equation}
A simple calculation(neglecting rest mass) gives
$$ dp= dE\Bigg[ 1+ \alpha_1 L_p E + ( {\frac {3}{2}} \alpha_2 - {\frac
{3}{8}}\alpha_{1}^{2} ) L_{p}^2 E^2 +(2\alpha_3 - \alpha_1 \alpha_2
+\frac{1}{4} \alpha_1^3) L_p^3 E^3+$$ $$( - \frac {5}{4} \alpha_1
\alpha_3 + \frac{15}{16}\alpha_1^2 \alpha_2 - \frac{5}{8}\alpha_2^2-
\frac{25}{128}\alpha_1^4)L_p^4 E^4 +$$
\begin{equation}
(-\frac{3}{2}\alpha_2\alpha_3 +\frac{9}{8}\alpha_1^2\alpha_3
+\frac{9}{8}\alpha_1\alpha_2^2 +\frac{21}{128}\alpha_1^5 -
\frac{45}{48}\alpha_2 \alpha_1^3)L_p^5 E^5\Bigg],
\end{equation}
then we find
$$ dE = dp \Bigg[1 - \alpha_1 L_p E +( - \frac {3}{2} \alpha_2 + \frac
{11}{8} \alpha_1^2) L_p^2 E^2 + (4\alpha_1\alpha_2 -2\alpha_1^3
-2\alpha_3) L_p^3 E^3 +$$ $$ ( \frac {23}{8} \alpha_2^2 +
\frac{21}{4} \alpha_1 \alpha_3 - \frac{137}{16} \alpha_1^2 \alpha_2
+ \frac{379}{128} \alpha_1^4) L_p^4 E^4 +$$
\begin{equation}
( \frac{15}{2} \alpha_2 \alpha_3 - \frac {97}{8} \alpha_1 \alpha_2^2
- \frac {89}{8} \alpha_1^2 \alpha_3 - \frac{565}{128} \alpha_1^5 +
\frac {801}{48} \alpha_1^3 \alpha_2 ) L_p^5 E^5 \Bigg].
\end{equation}
Within quantum field theory, the relation between particle
localization and its energy is given by $E\ge\frac{1}{\delta x}$,
where $\delta x$ is particle position uncertainty. Now it is obvious
that within MDRs, this relation should be modified. In a simple
analysis based on the familiar derivation of the relation
$E\ge\frac{1}{\delta x}$ [27], one can obtain the corresponding
generalized relation. This generalization is
$$ E \delta x \geq 1 + \frac {- \alpha_1 L_p}{\delta x} + \frac
{(\frac{11}{8}\alpha_1^2 - \frac {3}{2}\alpha_2)L_p^2}{\delta x^2} +
\frac {(4 \alpha_1 \alpha_2 - 2 \alpha_3 - 2 \alpha_1^3)
L_p^3}{\delta x^3}+$$ $$ \frac {(\frac{23}{8}\alpha_2^2 + \frac
{21}{4} \alpha_1 \alpha_3 - \frac{137}{16} \alpha_1^2 \alpha_2 +
\frac{379}{128} \alpha_1^4)L_p^4}{\delta x^4} +$$
\begin{equation}
\frac {(\frac{15}{2}\alpha_2 \alpha_3 - \frac{97}{8} \alpha_1
\alpha_2^2 - \frac{89}{8}\alpha_1^2 \alpha_3 -\frac {526}{128}
\alpha_1^5 + \frac{801}{48} \alpha_1^3 \alpha_2 )L_p^5}{\delta x^5}.
\end{equation}
In the same manner as previous section, the entropy of black hole
would be
$$S \simeq \frac{A}{4 L_p^2} + \frac{\alpha_1 \pi^{\frac{1}{2}}}{L_p}
A^{\frac{1}{2}} + \pi \Big(\frac{3}{2}\alpha_2 -
\frac{3}{8}\alpha_1^2\Big) \ln \frac{A}{L_{p}^{2}} - 4
\pi^{\frac{3}{2}} L_p \Big(-\alpha_1 \alpha_2 +\frac{1}{4}
\alpha_1^3 + 2 \alpha_3 \Big) A^{\frac{-1}{2}}$$
$$ - 4 \pi^2 L_p^2 \Big(-\frac{5}{4}\alpha_1 \alpha_3 - \frac{5}{8}
\alpha_2^2 + \frac{15}{16} \alpha_1^2 \alpha_2 - \frac{25}{128}
\alpha_1^4\Big) A^{-1}$$
\begin{equation}
\label{math:3.1} -\frac{16}{3} \pi^{\frac{5}{2}}L_p^3
\Big(\frac{9}{8} \alpha_1^2 \alpha_3 - \frac{45}{48}\alpha_1^3
\alpha_2 + \frac{9}{8} \alpha_1 \alpha_2^2 + \frac{21}{128}
\alpha_1^5 - \frac{3}{2} \alpha_2 \alpha_3\Big) A^{-\frac{3}{2}}
\end{equation}
It is easily seen that the entropy corrected by MDR has some terms
very different from string theory result. According to string
theory, the terms which include the half-odd power of $A$ or
$A^{-1}$ are not present in the entropy relation. Looking back to
our general form of MDR, (21), we see that if coefficients of the
odd power of energy in the modified dispersion relation were
vanishing($\alpha_1= \alpha_3= 0$), then unwanted terms in
entropy-area relation will disappear. Comparison between results of
MDR and string theory, suggests that in MDR, black hole entropy
should be
\begin{equation}
S \simeq \frac{A}{4 L_p^2} +\frac{3}{2}\pi\alpha_2 \ln
\frac{A}{L_{p}^{2}}+\frac{5}{2}\pi^2\alpha_{2}^2 \frac{L_p^2}{A}.
\end{equation}
We conclude that in MDR, all odd powers of energy should be omitted.
In other words, MDRs should contain only even power of energy. Using
equation (25), we find for temperature of black hole
\begin{equation}
T \simeq \frac {L_p^2}{8 \pi M} \Bigg[ 1 - \frac {\alpha_1 L_p}{2 M}
+ \frac{( \frac{11}{8} \alpha_1^2 -\frac{3}{2}\alpha_2)L_p^2}{4 M^2}
+ \frac{ ( 4\alpha_1 \alpha_2 - 2 \alpha_3 - 2 \alpha_1^3)L_p^3}{8
M^3}\Bigg ]
\end{equation}
Naturally, the presence of the even powers of the $\frac{1}{M}$
which are not present in string theory mass-temperature relation
(5), is due to $\alpha_i$ , where $i$ is odd. When we set
$\alpha_{i}=0$ for
all odd $i$, we find usual string theory result.\\
Now we answer the following question: what is the relation between
results of GUP and MDR?  First we consider corresponding relations
for entropy. These are equations (18) and (26),
$$
S \simeq \frac{A}{4 L_p^2} - \pi \alpha \ln \frac{A}{L_p^2}- 4
\pi^2(\alpha^2-\alpha^{''}) \frac{L_p^2}{A}\quad\quad
GUP\,\,Result,$$
$$ S \simeq \frac{A}{4 L_p^2} +\frac{3}{2}\pi\alpha_2 \ln
\frac{A}{L_{p}^{2}}+\frac{5}{2}\pi^2\alpha_{2}^2
\frac{L_p^2}{A}\quad\quad  MDR \,\, Result.$$\\
If we require these two results be consistent, we should have, for
example, $\alpha=-\frac{3}{2}\alpha_{2}$ and $\alpha''=
\frac{23}{8}\alpha_{2}^{2}$. This arguments show that actually GUP
and MDRs are not independent concepts. Since $\alpha$, $\alpha''$,
and ... are quantum gravity model dependent parameters, it seems
that in ultimate theory of quantum gravity, GUP and MDRs may be
equivalent concepts. Now, using string theory entropy-area relation,
(4), we see that $\rho=-\pi \alpha$ and $\beta=-4\pi^{2}(\alpha^2 -
\alpha'')$ for GUP-String theory correspondence, and
$\rho=\frac{3}{2}\pi \alpha_{2}$ and
$\beta=\frac{5}{2}\pi^{2}\alpha_{2}^{2}$ for MDR-string theory
correspondence.\\
Note that we have considered only a few terms of GUP and/or MDRs for
rest of our calculations, but considering more generalized form of
GUP and MDRs do not change our results regarding the form of GUP
and/or MDRs.

\section{Summary}
In this paper we have compared GUP and MDRs  quantum corrections of
black hole thermodynamics with more reliable string theory results.
Our comparison suggests that
\begin{itemize}
\item
In GUP, only even power of Planck length(or equivalently, only odd
power of $\frac{1}{\delta x}$) should be present.
\item
In MDRs, only even power of energy should be present.
\item
GUP and MDRs are not independent. It seems that they could be
equivalent concept in ultimate quantum gravity theory.
\item
Constraints on the form of GUP and/or MDRs may reflect inherent
features of underlying quantum gravity theory.
\end{itemize}
One may argue that our conclusions regarding GUP and/or MDRs
functional form, are not general since we have considered only a few
terms in GUP and/or MDRs. Actually  calculations based on more terms
in GUP and/or MDRs  support our results. This is reasonable at least
on symmetry grounds. Note that our arguments are based on the
assumption that today, string theory and loop quantum gravity
results are more reliable than other alternatives of quantum
gravity.

\end{document}